\documentclass[a4paper]{spicawStyle}
\usepackage{graphicx,natbib}
\pdfoutput=1

\begin{document}

\title{The Sizes of Kuiper Belt Objects}

\author{P. Lacerda\inst{1,2} } 

\institute{
Queen's University, Belfast BT7 1NN, United Kingdom.
\and 
Newton Fellow.}

\maketitle 

\begin{abstract}

One of the most fundamental problems in the study of Kuiper belt objects (KBOs)
is to know their true physical size. Without knowledge of their albedos we are
not able to distinguish large and dark from small and bright KBOs.
\emph{Spitzer} produced rough estimates of the sizes and albedos of about 20
KBOs, and the \emph{Herschel} space telescope will improve on those initial
measurements by extending the sample to the $\sim$150 brightest KBOs.
\emph{SPICA}'s higher sensitivity instruments should allow us not only
to broaden the sample to smaller KBOs but also to achieve a statistically
significant sample of KBO thermal light curves (\emph{Herschel} will measure
only six objects). A large sample covering a broad range of sizes will be key
to identify meaningful correlations between size and other physical and surface
properties that constrain the processes of formation and evolution of the solar
system.


\end{abstract}

\section{Introduction}

In this paper I discuss the importance of the upcoming \emph{`Space Infrared
Telescope for Cosmology \& Astrophysics'} (hereafter, \emph{SPICA}) for the
study of the icy small bodies of the outer solar system. Here, I focus on the
study of Kuiper belt objects (KBOs) but the same ideas can be applied to any
atmosphereless bodies. I will mainly discuss how \emph{SPICA} can help us to
measure the sizes of KBOs, and how that leads to more accurate estimates of the
size distribution and total mass of KBOs. I will also mention how this new
infrared space telescope might probe the rotational properties, chemical
composition and thermophysical parameters of those icy bodies. Other uses of a
space-based infrared telescope for solar system studies are detailed elsewhere,
in the context of \emph{SPICA}'s precursor observatory, \emph{Herschel}
\citep{2009EAS....34..133L}. In \S2 to \S4 I begin by summarising what we know
about KBOs and why their study is interesting and important.  Although many
accounts exist elsewhere I believe these proceedings should include a broad
overview of the subject. In \S5 and \S6 I discuss how \emph{SPICA} may
contribute to the study of KBOs.

\section{Our Solar System: the 1980's versus now}

Just a couple of decades ago our understanding of the solar system was quite
different from what it is today. Scientific interest lay mainly with the nine
planets: six (Mercury to Saturn) already known to the Greeks, another two
(Uranus and Neptune) discovered in the 18th and 19th centuries and a very
peculiar ninth planet (Pluto) discovered in 1930.  The small rocky planets, all
within 1.5 astronomical units (AU) from the Sun, stood in sharp contrast with
the outer gaseous giants extending out to 30 AU.  More intriguing was Pluto, a
moon-sized world on an eccentric and very inclined orbit beyond Neptune. Pluto
did not seem to fit in with the rest and was oddly isolated given its small
size.  Comets, asteroids, and even some planetary moons had a mysterious
character to them and were for the most part not well understood.

This view of the solar system has changed dramatically in the last 20 years or
so, mainly following the discovery of the Kuiper belt and of discs and
planetary systems around other stars. The focus of planetary science has moved
to the smaller bodies of the solar system as most of the interesting results
and paradigm-shifting discoveries have come from the study of their properties.
We begin to understand how the planets formed and evolved, how different
families of small bodies relate to one another, and how our solar system fits
in the larger picture of what we are finding in extrasolar planetary systems.

\section{The Kuiper Belt}


The Kuiper belt was identified in 1992 with the discovery of 1992 QB$_1$
\citep{1993Natur.362..730J}. Since then more than 1000 KBOs have been
discovered in the region roughly from 30 to 50 AU. The \emph{known} KBOs range
from about 25 to 2500 km in diameter
\citep{2004AJ....128.1364B,2008ssbn.book..335B} but numerous smaller objects
are believed to exist down to the micrometer-sized dust grains that have been
detected by Voyager 1 and 2 \citep{1997GeoRL..24.3125G}. Larger bodies could
also exist and remain undetected. The Kuiper belt is the solar system analogue
to the debris discs found around other stars \citep{2008ARA&A..46..339W}.

Kuiper belt objects provide us with probably the best picture of what the
planetesimals that formed the planets might have looked like. Dynamically, most
KBO orbits are stable against gravitational perturbations from the giant
planets on Gyr timescales. Chemically, the low temperatures found at such large
heliocentric distance ensure that KBOs preserve significant volatile content
from the protosolar nebula.  KBOs display the largest spread in surface colours
of all objects in the solar system \citep{2002AJ....123.1039J}. The origin of
the diversity is unknown but could reflect dynamical mixing resulting from
planetary migration \citep{2005Natur.435..459T}.  Physically, as the largest
($D>1000$ km) remnants of the planetesimals that formed the planets, KBOs
retain valuable information about the size, density and angular momentum
distributions at the poorly understood epoch of accretion.  The study of KBOs
can thus provide unique clues about the formation of solar system bodies.

\begin{figure}[ht]
  \begin{center}
    \includegraphics[width=8.5 cm]{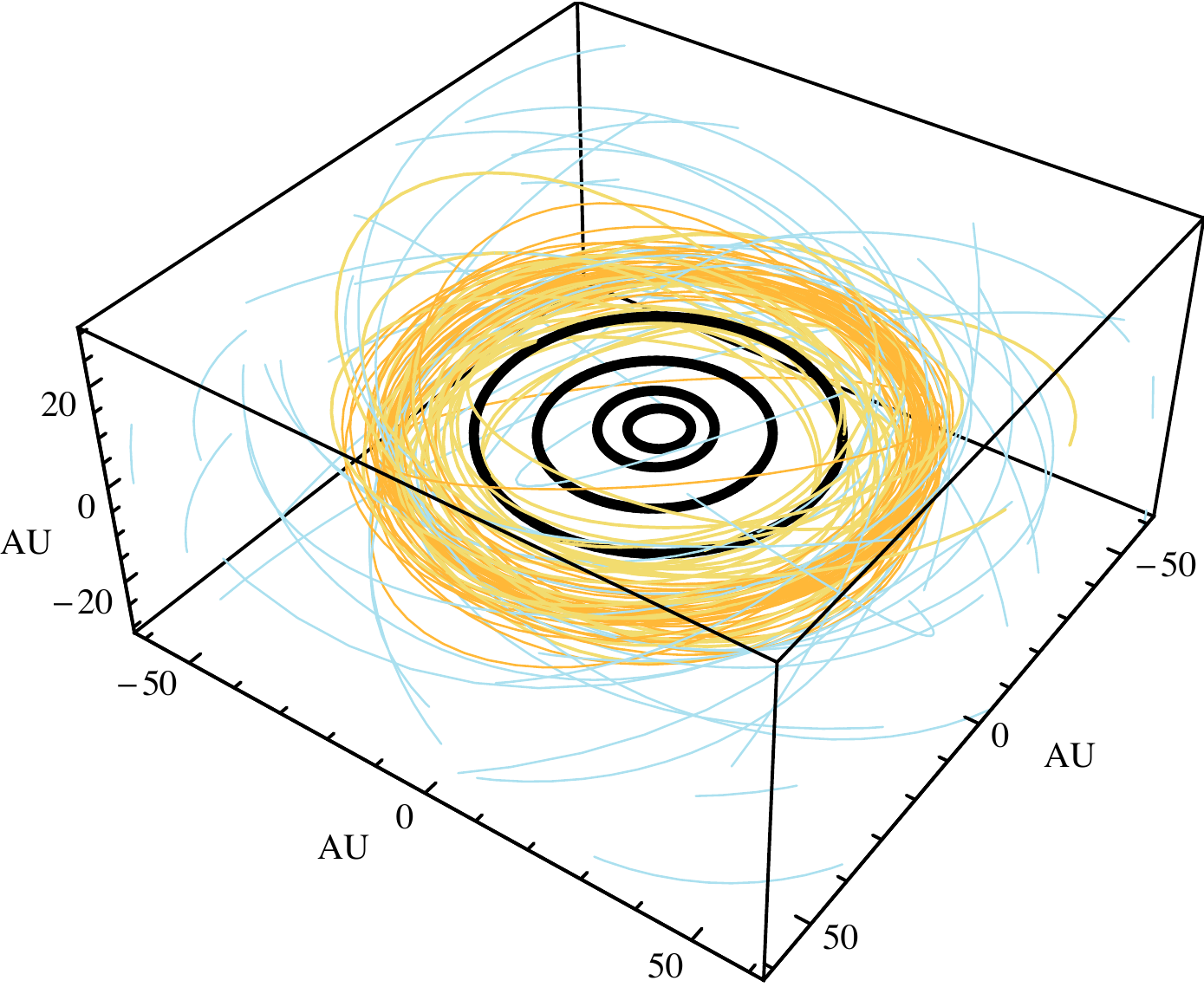}
  \end{center}
  \caption{Three-dimensional representation of the Kuiper belt.  Roughly 10\%
  of known orbits are plotted. The colours match those in Fig.\
  \ref{xxx20091029b_fig:fig2}. The thick black lines indicate the orbits of
  Jupiter, Saturn, Uranus and Neptune. Centaurs are not shown for clarity.}  
\label{xxx20091029b_fig:fig1}
\end{figure}

\begin{figure}[t]
  \begin{center}
    \includegraphics[width=8.5 cm]{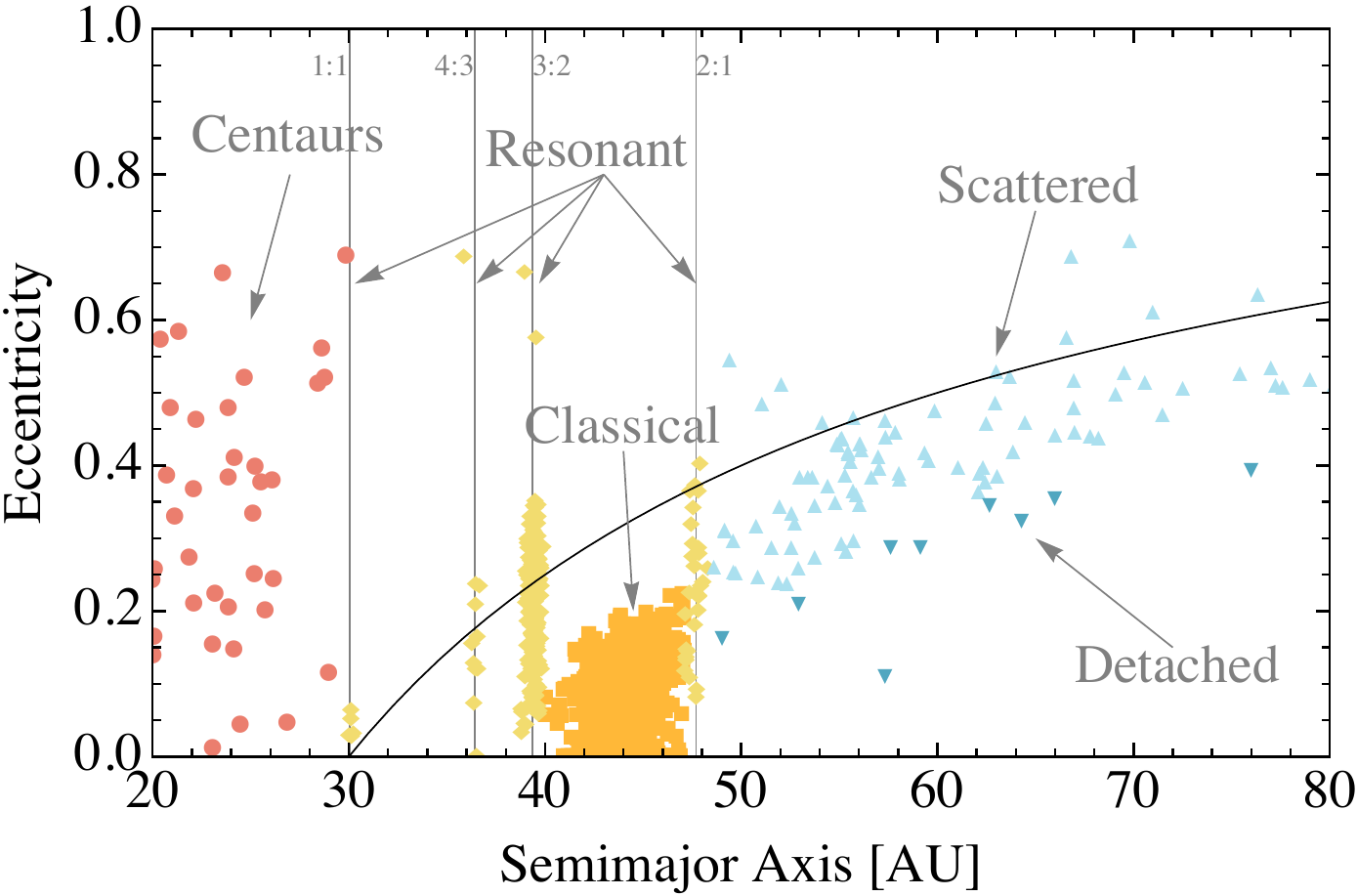}
  \end{center}
  \caption{Orbital structure of Kuiper belt objects. Grey vertical lines
  indicate mean motion resonances with planet Neptune. Objects plotted beyond
  30 AU and above the black solid curve cross the orbit of Neptune.}  
\label{xxx20091029b_fig:fig2}
\end{figure}

Kuiper belt orbits form a thick disc (7\% of known KBOs have inclinations
$i>30\degr$) beyond Neptune (Fig.\ \ref{xxx20091029b_fig:fig1}) but the orbital
distribution within that disc is not random. Most orbits fall in one of four
dynamical classes (Fig.\ \ref{xxx20091029b_fig:fig2}). 

\emph{Resonant KBOs} lie in mean motion resonances with Neptune. Because of
their resonant character they avoid close encounters with the massive planet
and are stable over the age of the solar system.  Pluto lies in the 3:2
resonance at roughly 39 AU from the Sun and so other KBOs in the same resonance
are often called Plutinos.

\emph{Classical KBOs} are the dynamically quintessential objects.  They have
relatively low eccentricity and low inclination orbits between the 3:2 and the
2:1 resonances at $\sim39$ AU and $\sim48$ AU. They are called classical
because their orbits most closely match what would be expected for a cold
dynamically stable disc. The classical belt turned out to have a broader
inclination distribution than first expected. 1992 QB$_1$, the first object
discovered in the Kuiper belt, is a classical KBO.

\emph{Scattered KBOs} are objects that interact strongly with Neptune near
perihelion and are thus being scattered by the giant planet. It is generally
believed that scattered KBOs originate in slightly unstable regions within the
Resonants and Classicals. Once in the scattered population these KBOs can be
thrown into planet crossing, \emph{Centaur}-type orbits which eventually feed
the Jupiter family comet population \citep{1997Sci...276.1670D}. This scenario
is not unique, though \citep{2008ApJ...687..714V}.  Centaurs are a population
closely related to KBOs. They have giant-planet-crossing orbits unstable on Myr
timescales. Possible end-states for Centaurs are collision with a giant planet,
or ejection from the solar system. Some return to the Scattered Kuiper belt and
a few end up (or spend part of their lifetime) as Jupiter family comets
\citep{2009Icar..203..155B}.

\emph{Detached KBOs} have perihelia beyond 40 AU indicating that Neptune had
little influence in their orbital evolution. They may have been emplaced at an
earlier epoch by the pull of a star passing close to the Sun or by past
gravitational perturbations by an unseen (or since ejected) planet beyond
\citep{2008AJ....135.1161L}.

\begin{table*}[bht]
  \caption{A selection of known KBOs and their properties.  Listed are object
  name, followed by approximate values for diametre, axis ratio (shape), spin
  period, light curve range, bulk density, surface albedo, surface composition,
  and orbital parametres. Question marks indicate unknown or poorly constrained
  entries.}
  \label{xxx20091029b_tab:tab1}
  \begin{center}
    \leavevmode
    \footnotesize
    \begin{tabular}[h]{lccrrcccccr}
      \hline \\[-5pt]
      Object          & D [km]& $a/b$ & $P$ [hr] & $\Delta m$ [mag]& $\rho$ [kg~m$^{-3}$] & Albedo  & Surface Comp.    &$a$ [AU]  & $e$  &$i$ [$^\circ$] \\[+5pt]
      \hline                                                                                                                                              \\[-10pt]
      Eris            & 2400  & 1.0  & ?        & $<0.01$          & 2300                  &  0.85   &CH$_4$            & 67.8     & 0.44 & 44.0          \\
      Pluto           & 2290  & 1.0  & 153.2    & 0.33             & 2000                  &  0.60   &CH$_4$, CO, N$_2$ & 39.6     & 0.24 & 17.1          \\
      Haumea          & 1500  & 1.2  & 3.9      & 0.29             & 2600                  &  0.70   &H$_2$O            & 43.2     & 0.19 & 28.2          \\
      Quaoar          & 1250  & 1.1  & 17.7     & 0.13             & ?                    &  0.09   &H$_2$O            & 43.1     & 0.04 & 8.0           \\
      Varuna          & 1000  & 1.3  & 6.3      & 0.42             & 1000                  &  0.07   &H$_2$O?           & 42.8     & 0.06 & 17.2          \\
      Huya            & 530   & 1.0  & ?        & $<0.06$          & ?                    &  0.06   &H$_2$O?           & 39.8     & 0.28 & 15.5          \\
      2000 GN$_{171}$ & 320   & 1.6  & 8.3      & 0.61             & 600                  &  0.06   &?                 & 39.7     & 0.29 & 10.8          \\
      2001 QG$_{298}$ & 230   & 2.9  & 13.8     & 1.14             & 600                  &  ?      &?                 & 39.6     & 0.20 & 6.5           \\
      \hline \\
      \end{tabular}
  \end{center}
\end{table*}

The orbital architecture of the Kuiper belt has provided many clues about the
past dynamical evolution of the solar system. For instance, we now know that
the giant planets did not form in their current locations but instead started
out in a much more compact configuration, from roughly 5 to 15 AU and then
migrated to their current positions. The stable resonances in the Kuiper belt
were populated as Neptune migrated outwards into a cold disc of planetesimals
by sweeping bodies as they moved along with the planet. Indeed, the most
encompassing model of the dynamical evolution of the solar system -- the
\emph{Nice} model \citep[named after the French city;][]{2005Natur.435..459T}
-- is largely designed to fit the observed properties of KBOs and small solar
system bodies in general. It fits, among other things, the global architecture
of solar system orbits, the migration of the planets by interaction with the
planetesimals, the enhanced lunar cratering record $\sim$3.8 Gyr ago known as
late heavy bombardment, the formation of the Oort cloud, the origin of Trojan
asteroids and the existence of hot (high-$i$) and cold (low-$i$) classical KBO
populations.
The strength and weakness of the \emph{Nice} model lies in its adaptability to
new observational discoveries.  New and more stringent observational
constraints should be sought to truly test the model.

The cumulative luminosity function of KBOs is well described by a power law,
$\log \Sigma=\alpha(m_R-m_0)$, with $\alpha\sim0.65$ and $m_0\sim23.5$
\citep{2001AJ....122..457T}. To translate luminosity into size requires knowing
the surface albedo. Assuming a uniform albedo for all KBOs and no relation
between heliocentric distance and size, the slope of the size distribution can
be inferred from that of the luminosity function as $q=4\alpha+1$
\citep{1995AJ....110.3082I}. However, the few known albedos (see examples in
Table~\ref{xxx20091029b_tab:tab1}) show that albedo is probably a strong
function of size and surface properties. Measuring and understanding the albedo
distribution is crucial if we are to constrain the size distribution and total
mass of the Kuiper belt; \emph{SPICA} is expected to play a major role in
achieving this goal.

Kuiper belt objects exhibit a tremendous diversity in surface colours,
unparalleled in the solar system \citep{1996AJ....112.2310L}. This probably
reflects significant chemical diversity, which is a puzzle given the small
range of temperatures in the 30 to 50 AU region. Spectroscopy, only possible
for the few brightest objects, shows that the largest KBOs are predominantly
coated in either methane ice (e.g. Eris, Pluto) or water ice (e.g. Haumea,
Quaoar).  Smaller objects are generally too faint for spectroscopic studies but
the few observed at sufficient S/N show mostly featureless spectra.

In summary, the Kuiper belt is now understood as a significant component of the
Sun's debris disc, as the most likely source of Centaurs and Jupiter family
comets, and possibly even of the Trojans, unusual planetary moons (e.g.
Triton, Phoebe) and the irregular satellites of the giant planets. The belt has
also provided a context for global models of the evolution of the solar system.

\section{Interesting KBOs}

{\em 1992 QB$_1$} was the first identified KBO. It is roughly 250 km in
diameter assuming a cometary albedo of 4\%. Its orbit is nearly circular and
has low inclination. 

{\em Varuna} was discovered on 28 Nov.\ 2000 by R.\ S.\ McMillan.  Being one of
the brightest known KBOs at the time, Varuna was intensely observed and
studied.  Combined sub-millimetre and optical observations were used to solve
the degeneracy between size and albedo \citep{2001Natur.411..446J} and estimate
Varuna's diameter ($\sim1000$ km) and albedo ($\sim0.07)$.  Light curve
observations revealed a rotationally deformed, fast-spinning object
\cite[$P_\mathrm{rot}=6.34$ hr;][]{2002AJ....123.2110J}.  Varuna is too large
to support significant topography and its overall shape is set by the balance
between gravitational and rotational accelerations. This property allows its
bulk density to be estimated
\citep[$\rho\sim1000$~kg~m$^{-3}$;][]{2007AJ....133.1393L}.  

{\em 1998 WW$_{31}$} was the first binary KBO to be discovered after
Pluto/Charon \citep{2002Natur.416..711V}.  Like most KBO binaries, this system
has nearly equal sized components.  Binaries are important because their total
mass can be estimated using Kepler's 3rd law and, if the size of the components
is known, their densities can be estimated.  

{\em 2001 QG$_{298}$} was found to display extremely large photometric
variability, $\Delta m=1.14\pm0.04$ mag \citep{2004AJ....127.3023S}. The large
$\Delta m$ combined with a relatively slow rotation, $P\sim13.8$ hr, suggest
this object is a contact binary. The shape of the components, as inferred from
the light curve, and the spin period imply a low bulk density of about
$\rho\sim650$~kg~m$^{-3}$ \citep{2007AJ....133.1393L}. Statistically, about
20\% of KBOs could display the extreme properties of QG$_{298}$, meaning that
many more await discovery.  The prospect of measuring densities from
contact-binary-type light curves is compelling.  

{\em Eris} was the first truly Pluto-sized KBO found
\citep{2006ApJ...643L..61B}; those are the only KBOs than can currently be
resolved by {\em HST}.  The possibility that Eris is larger than Pluto
intensified the planethood controversy.  Eris is covered in methane ice and has
a high albedo ($>80\%$) surface.  Its density is about 2300~kg~m$^{-3}$,
comparable to that of Pluto.  

{\em Haumea} is one of the strangest known KBOs. It spins extremely rapidly,
$P=3.9$~hr and, like Varuna, it is rotationally distorted into a triaxial shape
\citep{2006ApJ...639.1238R}. Unlike Pluto and Eris, Haumea is covered in almost
pure water ice but its high bulk density \citep[$\rho\sim2500$~kg~m$^{-3}$,
estimated in the same way as for Varuna; ][]{2007AJ....133.1393L} indicates
that it must have a rocky core and thus be differentiated.  A violent collision
has been proposed to explain Haumea's fast rotation, the fact that it has two
small, water-ice-rich moonlets, and the presence of half a dozen water-ice-rich
KBOs in its orbital vicinity. The collision would have happened more than 1 Gyr
ago, onto a proto-Haumea that was differentiated and had a thick water ice
mantle \citep{2007Natur.446..294B}.\looseness=-1

\section{The sizes of Kuiper belt objects}

Without knowing the surface albedo of a KBO it is not possible to tell large
and dark from small and bright objects. Ideally, the albedo can be measured by
combining visible and thermal infrared observations.  The reflected, visible
light is proportional to the KBO cross-section, $S$, and albedo, $A$, while the
thermal emission is proportional to $S\times(1-A)$, i.e. to the fraction of
light absorbed by the object which contributes to heating its surface. Visible
and thermal observations can thus be used to solve for $S$ and $A$. In
practice, other unknown parameters describing the spin state and orientation of
the KBO, and its surface roughness, emissivity, and thermal diffusivity (or
inertia) complicate the process and generally require more detailed
observations at different wavelengths.

\begin{figure}[ht]
  \begin{center}
    \includegraphics[width=8.5 cm]{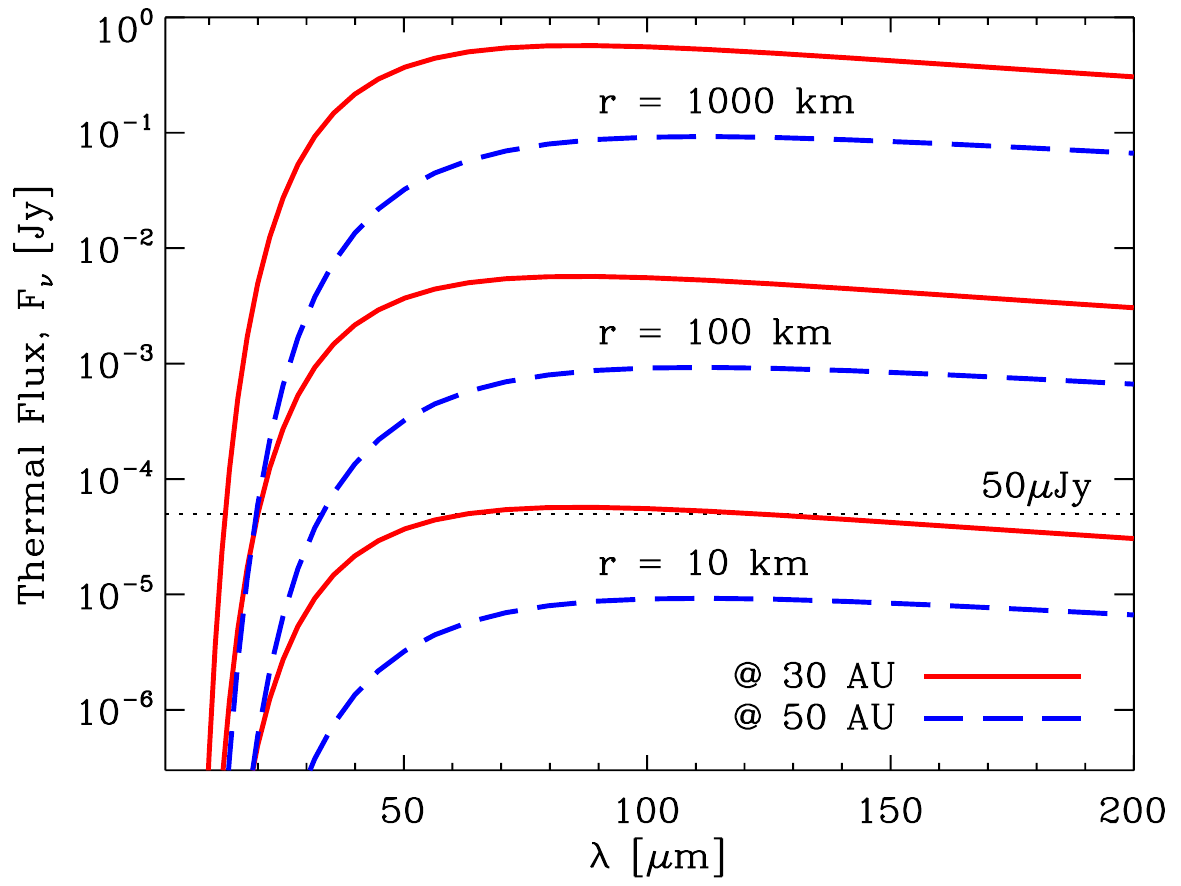}
  \end{center}
  \caption{Predicted flux density of KBOs of different radii at 30 and 50 AU
  from the Sun versus the wavelength coverage of \emph{SPICA}. The expected
  sensitivity of {\emph SPICA}, 50$\mu$m, is indicated by a horizontal dotted
  line.}  
\label{xxx20091029b_fig:fig3}
\end{figure}

Due to their great heliocentric distances, KBOs have very cold surfaces with
equilibrium temperatures around 40 to 50~K. Consequently, their thermal
emission peaks at 50 to 80~$\mu$m, well within \emph{SPICA}'s wavelength
coverage (Fig.\ \ref{xxx20091029b_fig:fig3}). But by far the most important
feature of \emph{SPICA} will be its enhanced sensitivity. Active cooling of the
mirror should bring the sensitivity down to $\sim$50~$\mu$Jy, nearly two orders
of magnitude better than \emph{Herschel}. \emph{SPICA} will be able to detect
objects almost as small as 10~km across, one order of magnitude smaller than
\emph{Herschel}, meaning that thousands more bodies will be accessible to the
new telescope. \emph{Herschel} will carry out a key project to observe
$\sim$140 large KBOs and Centaurs. \emph{SPICA} should vastly increase this
number and extend the sample to smaller bodies, approaching the sizes of comet
nuclei which could then be studied in their pristine state before they reach
the inner solar system.\looseness=-1

\begin{figure}[ht]
  \begin{center}
    \includegraphics[width=8.5 cm]{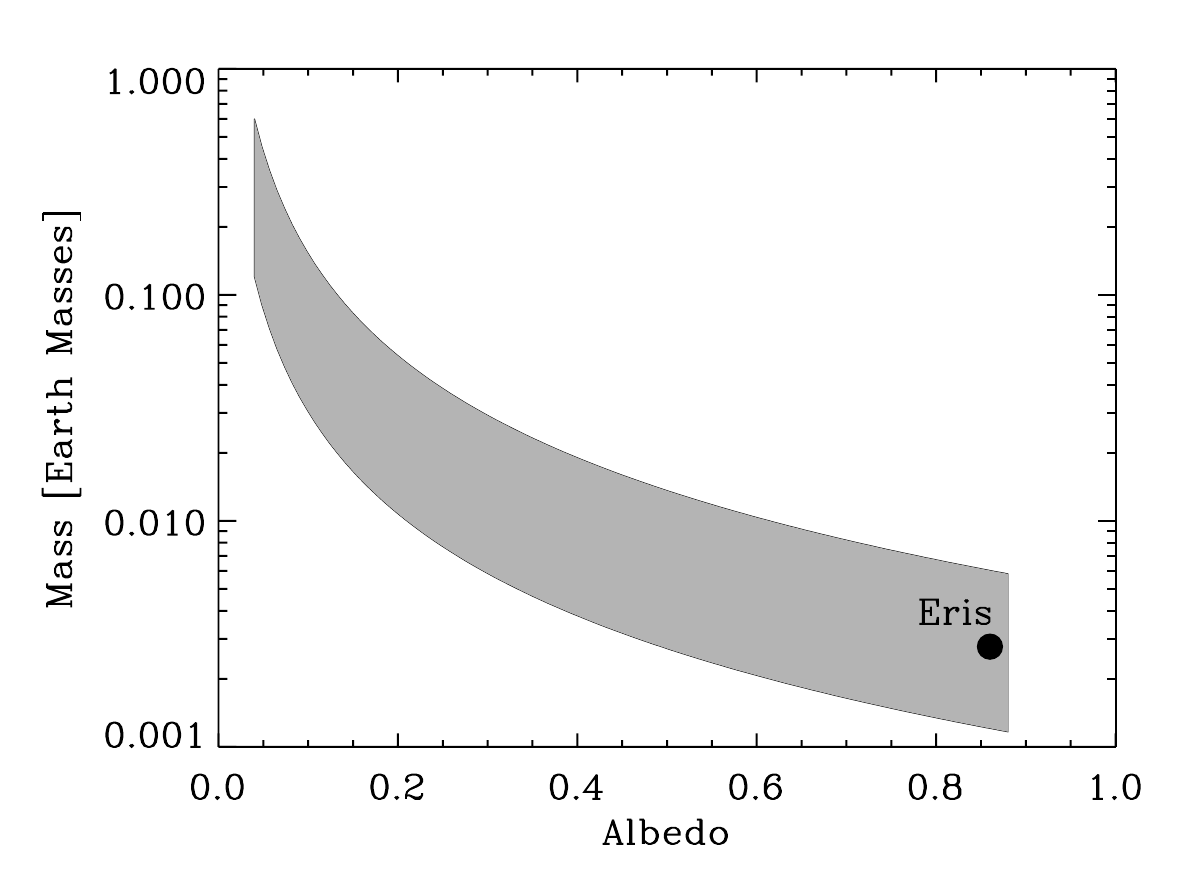}
  \end{center}
  \caption{Mass of large KBO Eris as a function of the assumed albedo.
  The current best estimate is plotted as a black dot. The vertical extent of
  the gray area reflects a range of plausible bulk densities, from 1 to 5~g
  cm$^{-3}$.}  
\label{xxx20091029b_fig:fig4}
\end{figure}

Another advantage of looking at the smaller bodies is to understand how albedo
varies with size.  Understanding the albedo distribution of KBOs is important
in many ways. For instance, estimating the size distribution and total mass of
the Kuiper belt from the observable luminosity function relies strongly on
assumptions about the albedo.  When derived from its brightness, the diameter
of a KBO varies as $A^{-1/2}$, where $A$ is the albedo, and its mass varies as
$A^{-3/2}$.  The mass estimate has the further complication that the densities
are also unknown. Plausible solar system densities range from
$\rho\sim500$~kg~m$^{-3}$ for comet nuclei, to $\rho\sim5000$~kg~m$^{-3}$ for
terrestrial planets. Figure \ref{xxx20091029b_fig:fig4} illustrates how
uncertainties in albedo and density can lead to differences of a few orders of
magnitude in the derived masses. The current estimated total mass of KBOs is
0.01 to 0.1~$M_\oplus$. Accretion models in the outer solar system require
10~$M_\oplus$ of material in the proto-Kuiper belt to explain the formation of
Pluto-sized objects \citep{1999AJ....118.1101K} in reasonable 10 to 100~Myr
timescales, and the \emph{Nice} model assumes 35~$M_\oplus$ to explain the
current orbital architecture of the giant planets \citep{2005Natur.435..459T}.
The \emph{Nice} model can also explain how $\sim$99\% of the initial mass has
been lost but the uncertainty in the current mass prevents it from offering a
solid constraint to the model.

For binary KBOs it becomes even more interesting to be able to measure their
physical size (cross-section). Binaries offer the opportunity of measuring mass
which can be translated into bulk density if we know the size.  Density is hard
to measure remotely but is very useful as first indicator of inner structure
and composition. Exactly how density depends on size may reveal whether the
former is more strongly influenced by composition or porosity. By
the time \emph{SPICA} begins operations, the next-generation space telescope
\emph{JWST} will presumably have identified many more binary KBOs suitable for
mass determination.

\section{Chemical and physical properties of KBOs}

Accurate albedos are also important for spectral modelling. Models by Hapke and
Shkuratov to investigate the composition and relative abundances of surface
materials can only be applied if the absolute reflectance is known.
\emph{SPICA}'s increased sensitivity will, for the first time, open the
far-infrared domain to spectroscopic studies of KBO surfaces. The far-infrared
is rich in diagnostic features diagnostic of silicates and ices expected to
incorporate KBOs. Amorphous and crystalline water ice can also be identified
through far-infrared spectral features at 44, 45 and 62~$\mu$m
\citep{1992ApJ...401..353M}. At low temperatures ice will form in the amorphous
state and at 40 to 50~K it should remain so for the age of the solar system.
However, whenever identified on the surfaces of the largest KBOs water ice
appears consistently crystalline
\citep{2004Natur.432..731J,2007ApJ...655.1172T}. Amorphous ice could exist in
subsurface layers though, or in smaller KBOs, as it is thought to drive
cometary activity of Centaurs beyond 5~AU by converting into the crystalline
phase as these objects move in from the cold Kuiper belt
\citep{2009AJ....137.4296J}.

The thermal emission of KBOs depends not only on their size and albedo, but
also on thermophysical properties of the surface such as emissivity and thermal
conductivity (or inertia), and on surface roughness. Multi-wavelength
measurements can be used to constrain these unknown parameters and provide
extra information about the surface of the distant KBOs. However, to accurately
model the temperature distribution across the surface of the KBO we need to
know its spin state (rotation period and spin orientation; see below).

\begin{figure}[ht]
  \begin{center}
    \includegraphics[width=8.0 cm]{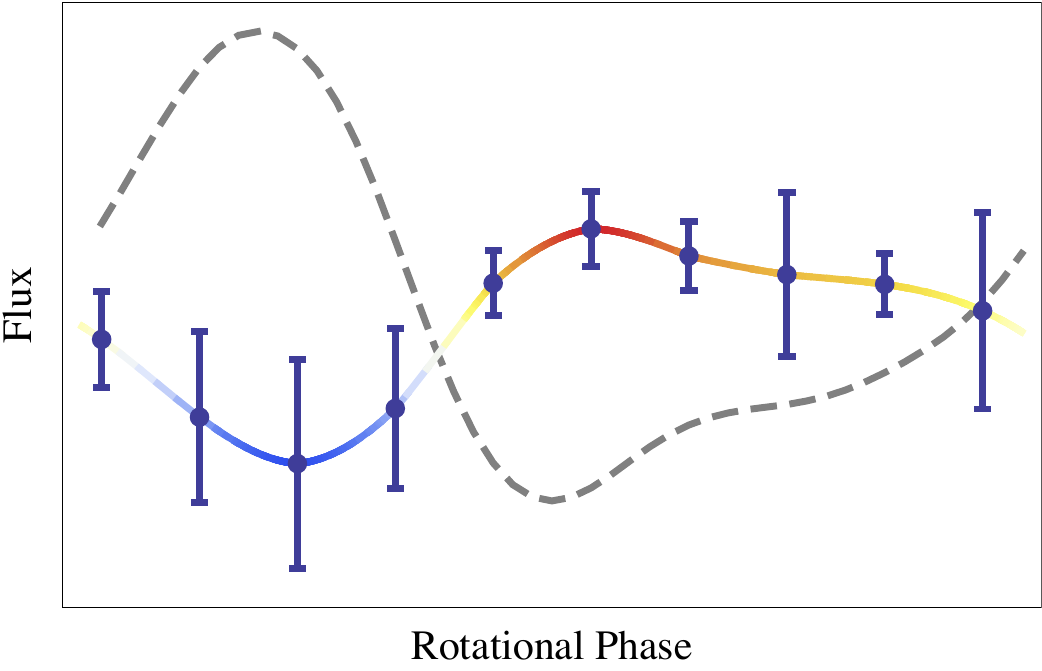}
  \end{center}
  \caption{Simulated visible (dashed grey) and thermal (points with error bars
  and solid line) light curves of a spherical object with less and more
  reflective patches across its surface. The more reflective areas are brighter
  in visible light but cooler and thus fainter at thermal wavelengths.
  Conversely, the optically darker patches absorb more solar radiation and
  appear warmer, hence brighter in the far-infrared. In summary, an optical
  light curve caused by albedo appears anti-correlated with its thermal
  counterpart. Deviations from perfect anti-correlation are due to thermal
  inertia  of the surface.}  
\label{xxx20091029b_fig:fig5}
\end{figure}

\begin{figure}[ht]
  \begin{center}
    \includegraphics[width=8.0 cm]{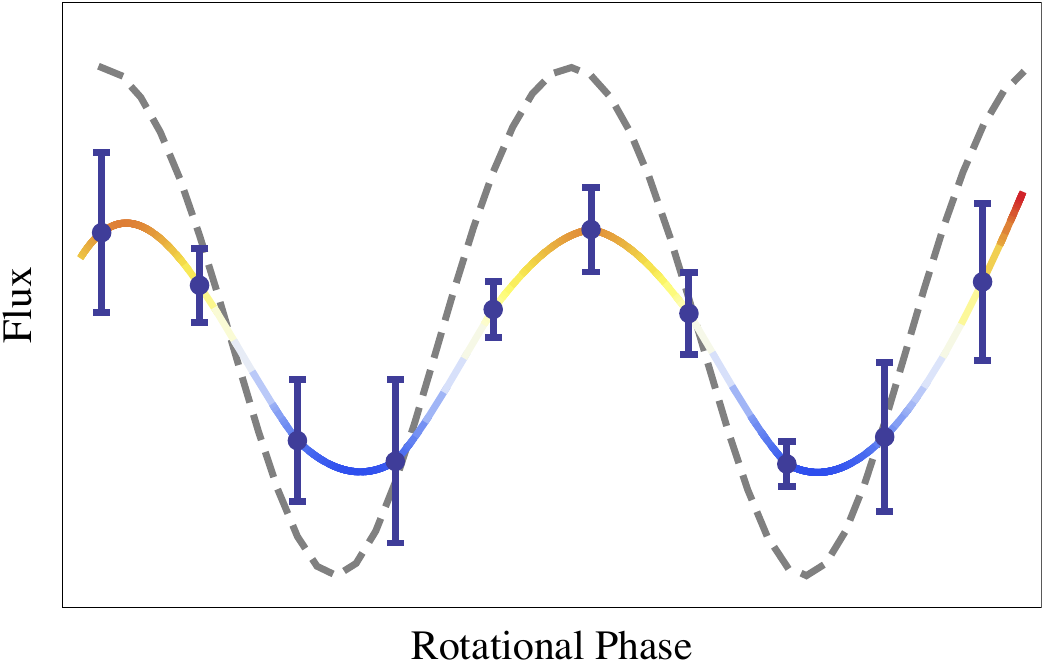}
  \end{center}
  \caption{Same as Fig.\ \ref{xxx20091029b_fig:fig5} but for an elongated
  object with a uniform surface. Here, the brightness modulation in the
  reflected light (dashed grey) is due to the varying apparent cross-section as
  the object rotates. The thermal light curve (points with error bars and solid
  line) follows the same pattern because the emitting cross-section is larger
  when the reflecting cross-section is larger.}  
\label{xxx20091029b_fig:fig6}
\end{figure}

Six of the brightest KBOs to be observed by \emph{Herschel} will be observed
repeatedly as they spin to obtain thermal emission curves. Given \emph{SPICA}'s
high sensitivity we will be able to extend this type of observation to many
more objects. By comparing the optical and thermal light curves we will be able
to establish with certainty whether the photometric variability is due to shape
or albedo spottiness. If the variability is due to albedo markings then
optically bright, high albedo regions will be fainter at thermal wavelengths
because they reflect more sunlight and remain cooler -- the optical and thermal
light curves appear uncorrelated (see Fig.\ \ref{xxx20091029b_fig:fig5}). If
due to shape, the body will be brighter at optical and thermal wavelengths at
roughly the same time (Fig.\ \ref{xxx20091029b_fig:fig6}). Breaking this
degeneracy between albedo and shape is important for studies of the shape
distribution and angular momentum content of KBOs
\citep{2003Icar..161..174L,2006AJ....131.2314L}. 

Time-resolved thermal observations may also constrain the orientation of KBO
spin axes w.r.t. the line of sight. The closer the spin vector lies to the line
of sight the warmer the object will be, on average, as a large fraction of its
surface close to the pole is continuously exposed to sunlight.  Modulations due
to shape will also be smaller than if the object is seen equator-on.
\emph{SPICA} will extend these techniques to a much larger sample and for those
six objects previously studies by \emph{Herschel} the observations about a
decade apart may help constrain the longitude of the spin pole as well. The
alignment of KBO spin poles can test models of planetesimal formation and
collisional evolution \citep{2005PhDT........21L,2009arXiv0910.1524J}.

\section{Conclusions}

The enhanced sensitivity of the upcoming \emph{`Space Infrared Telescope for
Cosmology \& Astrophysics'} (\emph{SPICA}) will play a key role in the study of
Kuiper belt objects.  Multi-wavelength thermal observations will enable us to
measure the albedos and sizes of several hundreds of bodies and to investigate
their surface composition and thermophysical properties. The high sensitivity
-- two orders of magnitude better than \emph{Herschel} -- will allow the
detection of KBOs as small as a few tens of kilometres across, very close to
the sizes of cometary nuclei. Sizes measurements of binary Kuiper belt objects
will lead to estimates of their bulk density, a first step into the interior
structure of these bodies. Time resolved thermal observations of a statistical
ensemble of KBOs will help break the degeneracy between albedo variability and
shape as the cause for observed light curves, which is important for assessing
the shape and angular momentum distributions of Kuiper belt objects.  By
extending the far-infrared domain to potentially hundreds of Kuiper belt
objects and associated families, \emph{SPICA} should allow us to identify
meaningful correlations and patterns between orbital, physical and chemical
properties and thus help us probe the formation and evolutionary processes that
operated at the early stages of our solar system.

\begin{acknowledgements}

I am grateful to David Jewitt for comments on the manuscript and to the Royal
Society for the support of a Newton Fellowship.\looseness=-1

\end{acknowledgements}

%
%


\begin{thebibliography}{}

\bibitem[Bailey \& Malhotra(2009)]{2009Icar..203..155B} Bailey, B.L., \&
Malhotra, R.\ 2009, Icarus 203, 155.


\bibitem[Bernstein et al.(2004)]{2004AJ....128.1364B} Bernstein, G.M., et al.\
2004, AJ 128, 1364.

\bibitem[Brown(2008)]{2008ssbn.book..335B} Brown, M.E.\ 2008, The Solar System
Beyond Neptune (The University of Arizona Press), 335.


\bibitem[Brown et al.(2006)]{2006ApJ...643L..61B} Brown, M.E., et al.\ 2006,
ApJL 643, L61.


\bibitem[Brown et al.(2007)]{2007Natur.446..294B} Brown, M.E., et al.\ 2007,
Nature 446, 294.

\bibitem[Duncan \& Levison(1997)]{1997Sci...276.1670D} Duncan, M.J., \&
Levison, H.F.\ 1997, Science 276, 1670.


\bibitem[Gurnett et al.(1997)]{1997GeoRL..24.3125G} Gurnett, D.A., et al.\
1997, GeoRL 24, 3125.

\bibitem[Irwin et al.(1995)]{1995AJ....110.3082I} Irwin, M.,
Tremaine, S., \& Zytkow, A.~N.\ 1995, AJ 110, 3082.

\bibitem[Jewitt(2002)]{2002AJ....123.1039J} Jewitt, D.C.\ 2002, AJ 123, 1039.

\bibitem[Jewitt(2009)]{2009AJ....137.4296J} Jewitt, D.\ 2009, AJ 137, 4296.

\bibitem[Jewitt et al.(2001)]{2001Natur.411..446J} Jewitt, D., Aussel, H., \&
Evans, A.\ 2001, Nature 411, 446.

\bibitem[Jewitt \& Luu(1993)]{1993Natur.362..730J} Jewitt, D., \& Luu, J.\ 1993,
Nature 362, 730.

\bibitem[Jewitt \& Luu(2004)]{2004Natur.432..731J} Jewitt, D.~C., \& Luu, J.\
2004, Nature 432, 731.

\bibitem[Jewitt \& Sheppard(2002)]{2002AJ....123.2110J} Jewitt, D.C., \&
Sheppard, S.S.\ 2002, AJ 123, 2110.

\bibitem[Johansen \& Lacerda(2009)]{2009arXiv0910.1524J} Johansen, A., \&
Lacerda, P.\ 2009, MNRAS, in press.

\bibitem[Kenyon \& Luu(1999)]{1999AJ....118.1101K} Kenyon, S.J., \& Luu,
J.X.\ 1999, AJ 118, 1101.

\bibitem[Lacerda(2005)]{2005PhDT........21L} Lacerda, P.\ 2005, Ph.D.~Thesis.

\bibitem[Lacerda \& Luu(2003)]{2003Icar..161..174L} Lacerda, P., \& Luu, J.\
2003, Icarus 161, 174.

\bibitem[Lacerda \& Luu(2006)]{2006AJ....131.2314L} Lacerda, P., \& Luu, J.\
2006, AJ 131, 2314.

\bibitem[Lacerda \& Jewitt(2007)]{2007AJ....133.1393L} Lacerda, P., \& Jewitt,
D.C.\ 2007, AJ 133, 1393.

\bibitem[Lellouch(2009)]{2009EAS....34..133L} Lellouch, E.\ 2009, EAS 
Publications Series 34, 133.

\bibitem[Luu \& Jewitt(1996)]{1996AJ....112.2310L} Luu, J., \& Jewitt, D.\
1996, AJ 112, 2310.

\bibitem[Lykawka \& Mukai(2008)]{2008AJ....135.1161L} Lykawka, P.~S., \& Mukai,
T.\ 2008, AJ 135, 1161.

\bibitem[Moore \& Hudson(1992)]{1992ApJ...401..353M} Moore, M.~H., \& Hudson,
R.~L.\ 1992, ApJ 401, 353.

\bibitem[Rabinowitz et al.(2006)]{2006ApJ...639.1238R} Rabinowitz, D.L., et al.\ 2006, ApJ 639, 1238.

\bibitem[Sheppard \& Jewitt(2004)]{2004AJ....127.3023S} Sheppard, S.S., \&
Jewitt, D.\ 2004, AJ 127, 3023.

\bibitem[Trujillo et al.(2001)]{2001AJ....122..457T} Trujillo, C.A.,
Jewitt, D.C., \& Luu, J.X.\ 2001, AJ 122, 457.

\bibitem[Trujillo et al.(2007)]{2007ApJ...655.1172T} Trujillo, C.A, el al.\
2007, ApJ 655, 1172.


\bibitem[Tsiganis et al.(2005)]{2005Natur.435..459T} Tsiganis, K., et al.\
2005, Nature 435, 459.

\bibitem[Veillet et al.(2002)]{2002Natur.416..711V} Veillet, C., et al.\ 
2002, Nature 416, 711.

\bibitem[Volk \& Malhotra(2008)]{2008ApJ...687..714V} Volk, K., \& Malhotra,
R.\ 2008, ApJ 687, 714.

\bibitem[Wyatt(2008)]{2008ARA&A..46..339W} Wyatt, M.C.\ 2008, ARA\&A 46, 339.

\end{thebibliography}
\end{document}